# Absorption mechanism of dopamine/DOPAC modified TiO$_2$ nanoparticles by time-dependent density functional theory calculations

Costanza Ronchi,[1,2,†] Federico Soria,[1†] Lorenzo Ferraro,[1] Silvana Botti,[2] Cristiana Di Valentin[1,*]

[1]Dipartimento di Scienza dei Materiali, Università di Milano Bicocca

Via R. Cozzi 55, 20125 Milano Italy

[2]Friedrich Schiller University Jena, Institut für Festkörpertheorie und -optik,

Max-Wien-Platz 107743 Jena, Germany

## Abstract

Donor-modified TiO$_2$ nanoparticles are interesting hybrid systems shifting the absorption edge of this semiconductor from the ultra-violet to the visible or infrared light spectrum, which is a benefit for several applications ranging from photochemistry, photocatalysis, photovoltaics, or photodynamic therapy. Here, we investigate the absorption properties of two catechol-like molecules, i.e. dopamine and DOPAC ligands, when anchored to a spherical anatase TiO$_2$ nanoparticle of realistic size (2.2 nm), by means of time-dependent density functional theory calculations. By the differential absorbance spectra with the bare nanoparticle, we show how it is possible to determine the injection mechanism. Since new low-energy absorption peaks are observed, we infer a direct charge transfer injection, which, unexpectedly, does not involve the lowest energy conduction band states. We also find that the more perpendicular the molecular benzene ring is to the surface, the more intense is the absorption, which suggests aiming at high molecular packing in the synthesis. Through a comparative investigation with a flat TiO$_2$ surface model, we unravel both the curvature and coverage effects.

[†] These authors have equally contributed to this work

[*] Corresponding author: cristiana.divalentin@unimib.it

## 1. INTRODUCTION

Titanium dioxide ($TiO_2$) nanoparticles have been effectively used in a wide range of applications that exploit their ability to convert light photons energy into chemical processes, including photocatalysis [1-4], photoelectrochemistry [5-6] photovoltaics [7-11] and even photochromic devices [12]. More recently, the possibility to exploit the photocatalytic properties of $TiO_2$ in nanomedicine has attracted the attention of researchers because of their possible application in the photodynamic therapy of tumors and for the photoinduced drug release [13].

Photoexcited hole and electron in $TiO_2$ nanoparticles (NPs) are capable of generating reactive oxygen species (ROS), such as $OH^{\bullet}$, $O_2^{2-}$ and $H_2O_2$, under UV light [14,15], which can drive several chemical reactions due to their high redox activity and can effectively induce cell death [16,17]. The fact that the $TiO_2$ NPs are cytotoxic only once they are light-activated is crucial to selectively kill only the targeted sick cells, limiting the side effects on the rest of the body and making them good candidates for cancer treatment [18].

The photocatalytic activity of $TiO_2$ is highly dependent on several parameters, including surface area, crystalline phase, and single crystallinity [19,20]. The anatase crystalline phase of $TiO_2$ presents superior photocatalytic properties with respect to the other phases, such as rutile [21]. The most stable anatase NP shape is a decahedron exposing two facets: (101) and (001). However, spherically shaped NPs are obtained in conditions of excessive dilution [13], which, presenting a higher number of low coordinated Ti atoms on the surface [22,23] are better suited for the surface functionalization [24-27] and, thus, more interesting for biomedical applications[24], such as drug transport or imaging [25].

Surface coating of $TiO_2$ nanoparticles is used to avoid NP agglomeration and improve biocompatibility. Surface functionalization can also be used to narrow the semiconductor band gap modifying its range of absorption [28,29]. The surface binding of organic molecules has significant impact on charge separation, transport, and recombination processes. Once functionalized, $TiO_2$ nanostructures develop the ability to harvest a major portion of the solar spectrum, becoming suitable for photon energy conversion, both for nanomedical [29] and environmental applications [30].

When the surface-functionalized $TiO_2$ nanosystems are properly irradiated, an electron is excited from the molecular moiety, which works as an electron donor or reductant, to the semiconducting oxide, which is the electron acceptor or oxidant. The electron transfer from the HOMO of an adsorbed molecule to the conduction band (CB) of a semiconductor has been modelled in the literature through two different limit mechanisms [31], depending on the electron coupling between the different components of the system (**Scheme 1**). On one hand, for weakly coupled systems, indirect injection, usually described as type I mechanism, proceeds in two steps [31]: one electron is initially promoted to the LUMO of the molecule and later transferred to the solid. On the other hand, strongly-coupled systems are better represented through a direct or type II mechanism, where an electron from the HOMO of the molecule is injected in the CB of the semiconductor in one direct step [32-36]. This mechanism of charge transfer (similar to a ligand-to-metal charge transfer or LMCT in transition metal complexes) is often referred to as ultra-fast direct injection [32,33] and is evidenced by a new band in the absorption spectrum that arises when the molecule is attached to the semiconductor surface [34-36]. According to this model, the transition dipole moment of the electronic excitation, which indicates the probability to observe such excitation, is directly proportional to the coupling between the states of the semiconductor and of the molecules. According to this selection rule, the charge transfer does not necessarily involve the lowest states at the CBM but it involves those states in the CB that are mostly coupled with the unoccupied molecular states.

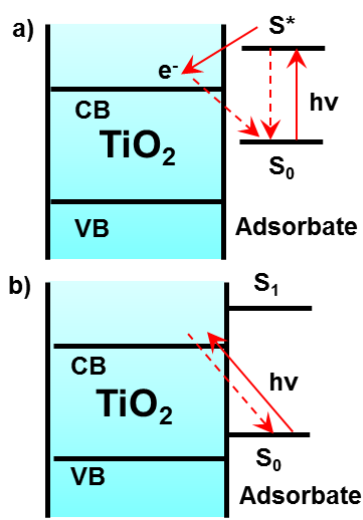

**Scheme 1**: Schematic representation of the two different injection mechanisms upon irradiation of an adsorbed molecule on a $TiO_2$ surface: a) type I or indirect mechanism. b) type II or direct mechanism.

From a theoretical point of view [37], time-dependent density functional theory (TDDFT) [38] has been used to describe the optical properties of $TiO_2$-modified systems [33,36,39-46]. TDDFT, which extends density functional theory into the time domain, is capable of describing excited states. System composed of $TiO_2$ modified by organic dyes were modelled through small clusters, $(TiO_2)_9$, by Sánchez-de-Armas et al.[36]. Through the comparison of the spectra for free and adsorbed dyes on $TiO_2$, together with the analysis of the molecular orbitals, the authors proposed either direct or indirect mechanism for the electron injection and even suggested the existence of possible intermediate mechanisms, depending on the energy difference between the HOMO and the LUMO of the molecule. In addition, a TDDFT study of the interaction of catechol and dopamine with $(TiO_2)_n$ clusters (n = 2, 4, 6) has also been performed [39], where the authors observed a red-shift of absorption for the molecule-cluster systems in comparison with that of isolated molecules and clusters. Moreover, the chelated adsorption mode was found to present bands at lower energies than the other modes [39]. Moving to bigger $TiO_2$ clusters, a $(TiO_2)_{38}$ exposing the majority (101) surface was extensively used by De Angelis et al. with Ru-based dyes [33,40], perylene, [31], [Fe(CN)$_6$] [41] and phenothiazine [42] whereas Zhao et al. used the same cluster to study a series of novel metal-free dyes based on the C217 [43]. Sanchez's group has used a cluster of 90 units of $TiO_2$ to study charge injection mechanism of different dyes in a coupled dye−$TiO_2$ NP using time-dependent self-consistent density functional tight-binding (TD-DFTB) [34,35]. They clearly observed the appearance of a new absorption band only for those dyes undergoing a direct photoinjection mechanism (type-II). Through periodic calculations, both the optical properties of dye-sensitized anatase $TiO_2$ flat surfaces and nanowires were also simulated by TDDFT, which allowed to highlight the role played by explicit solvent molecules and by dynamical effects [44-46].

Catechol derivatives bind extremely strong to the $TiO_2$ surface and show a high stabilization through the bidentate binding [28,29,47]. Between these catechol-like molecules, dopamine and DOPAC are two interesting cases, because they have double functionality: two hydroxyl groups on one side and one ethyl-amino functionality (dopamine) or a carboxyl group (DOPAC) on the other side, which can interact with other molecules, such as with the solvent or with drugs [48-50].

When a nanoparticle is modified by dopamine or DOPAC molecules, an absorption shift from UV to the visible region is observed because new molecular states appear in the band gap. Dopamine-functionalized $TiO_2$ NPs (DOP-NPs) show a shift of the light absorption onset from 380 to 800 nm. An increase in the dopamine coverage enhances the absorbance at all wavelengths <800 nm but does not shift the absorption onset, leading to the effective band gap of dopamine-modified $TiO_2$ of 1.6 eV [28,29,50-53]. Surface-enhanced Raman spectroscopy (SER) measurements suggests an effective charge transfer mechanism for DOP-NPs [52] with an unexpected dependence on the dopamine coverage and on the size of the NP. Also, in the case of DOPAC-NPs, the Raman signal intensity was found to depend on the number of surface binding sites, electron density of the ligands and dipole moment [53].

Both, dopamine- and DOPAC-functionalized $TiO_2$ nanosystems are powerful tools for biomanipulation. Recently, dopamine-modified $TiO_2$ was used to label the NP with a fluorescent marker [54] and as a drug delivery system of doxorubicin [55], whereas DOPAC-modified $TiO_2$ NPs carrying antibodies were used for targeted cancer therapy [29] and carrying epidermal growth factor receptors for both cellular and subcellular delivery [56].

Although these catechol-like bifunctional linkers attached to metal oxides nanoparticles are commonly used in biomedical applications, as summarised above, atomistic details and optical characterization of these hybrid systems from a theoretical point of view are still missing.

In this work, we present a TDDFT investigation of dopamine- and DOPAC-modified $TiO_2$ NPs, using realistic spherical $TiO_2$ models of 700 atoms (diameter size of 2.2 nm) [22,23,27,57]. First, we calculate the optical absorption cross section using the time-propagation method in TDDFT for the separated components and, then, for one linker molecule attached to the spherical $TiO_2$ NP. On this basis, we derive the difference absorption spectrum, which allows determining the absorption mechanism in these systems. For comparison, we also present a parallel study on a simplified periodic flat anatase (101) $TiO_2$ surface models going from low up to full coverage of dopamine molecules. The method of choice for this is the standard gradient corrected PBE functional [58]. To prove its reliability, we have performed a preliminary validation, using a small cluster of six $TiO_2$ units, where we compare PBE results with those by a more sophisticated range separated hybrid functional (CAM-B3LYP) [59] and by adding the solvent effect through the PCM model [60], as presented in the Computational details below.

## 2. COMPUTATIONAL DETAILS

The calculations performed in this work are based on two levels of theory: density functional theory (DFT) and time-dependent density functional theory (TDDFT). The first has been employed for geometry optimization and electronic structure calculations, whereas the second for simulation of the optical properties.

For the DFT calculations, we used the CRYSTAL14 package [61], where the Kohn-Sham orbitals are expanded in Gaussian-type orbitals, with the PBE exchange-correlation functional [58]. The all-electron basis sets are Ti 86-4111(d41), O 8-4111(d1) for the oxygen atoms of $TiO_2$; H 5-111(p1), C 6-31111 (d1), O 8-41111 (d1) and N 6-311(d1) have been employed for hydrogen, carbon, oxygen and nitrogen of the adsorbed molecules. The cut-off limits in the evaluation of Coulomb and exchange series/sums appearing in the SCF equation were set to $10^{-7}$ for Coulomb overlap tolerance, $10^{-7}$ for Coulomb penetration tolerance, $10^{-7}$ for exchange overlap tolerance, $10^{-7}$ for exchange pseudo-overlap in the direct space, and $10^{-14}$ for exchange pseudo-overlap. The condition for the SCF convergence was set to $10^{-6}$ au on the total energy difference between two subsequent cycles. The equilibrium structure of isolated molecules, NP and NP+molecule complexes were determined by using a quasi-Newton algorithm with a BFGS Hessian updating scheme [62]. Geometry optimization was performed without any symmetry constraint; forces were relaxed to be less than $4.5 \times 10^{-4}$ au, and displacements to be less than $1.8 \times 10^{-3}$ au. Optimized

PBE geometries have been compared to B3LYP-D* ones including the dispersion correction [27]. The deviations in the bond distances are within 1-2%.

The anatase $TiO_2$ spherical nanoparticle (NP) model used in this work was built previously by our group through global optimization with a simulated annealing process at the self-consistent charge density functional tight binding SCC-DFTB level of theory [23]. The stoichiometry of the model is $(TiO_2)_{223}\cdot 10H_2O$ and it is characterized by an equivalent diameter of 2.2 nm. The nanoparticle has been treated as a large isolated molecule in vacuum without any periodic boundary condition.

The (101) anatase surface was modelled with a periodic slab of three tri-atomic $TiO_2$ layers. The bottom layer was kept fixed at the optimized bulk positions during the geometry optimization. Periodicity was considered only along the $[10\bar{1}]$ and [010] directions, while no periodic boundary conditions were imposed in the direction perpendicular to the surface. We used two supercell models: a 2×4 supercell model ($a$=15.2119 Å, $b$= 11.1501 Å, 144 atoms) already described in a previous work of our research group to represent a low coverage functionalization (0.25 monolayer (ML)) and 1×2 supercell model ($a$=7.6059 Å, $b$= 5.5750 Å, 36 atoms) to represent a full coverage functionalization (1 ML) [63]. The configurations have been optimized using a $k$-point mesh of 4 × 4 × 1 to ensure the convergence of the electronic structure. The total density of states (DOS) and projected density of states (PDOS) were computed with a finer k-point mesh of 30×30×1.

The total adsorption energy per molecule has been defined as:

$$\Delta E_{ads}^{mol} = (E_{slab+nmol} - [E_{slab} + n_{mol}E_{mol}])/n_{mol}$$

The optical response has been obtained by real time time-dependent DFT (RT-TDDFT) calculations in the Kohn-Sham scheme, using the pseudopotential real-space code Octopus [64]. Following a ground-state calculation with PBE exchange-correlation functional [58], spectra are computed using the explicit time-propagation method, in which the Kohn-Sham orbitals are propagated over a series of discrete steps in real time. The time-propagation method offers a number of advantages over the "summing-over-states" methods [65,66]. Calculation times for the explicit time-propagation methods scale approximately with $N^2$ in comparison with the $\sim N^5$ scaling of the summing-over-states methods, where N is the number of atoms being considered [67].

Optimized Norm-Conserving Vanderbilt PBE pseudopotentials have been used. The spacing of the real-space grid was set to 0.18 Å, a spherical box shape with a radius of 20 Å was used for the molecules, NP and NP+molecule complexes. The time step for the propagation was set at $7.9 \times 10^{-4}$ fs, whereas the overall propagation time was variated from 0.72 to 5.4 fs for the molecules and 6.58 fs (10 ℏ/hartree) was used for the NP and NP+molecule, respectively. The propagation was carried out by means of the approximated enforced time-reversal symmetry (AETRS) propagator, [68] as implemented in the Octopus code.

For the slab calculations, we arbitrarily chose one direction for the calculation of the optical response. We calculated the response to a perturbation, using a time step for the propagation of 2.5 × $10^{-4}$ fs and a propagation time of 5 fs, in the x direction according to the orientation shown in the respective figures. In the TDDFT calculations, we used a *k*-point mesh of 4 × 4 × 1 for 1×2 supercell model and a mesh of 2 × 2 × 1 for 2×4 supercell model.

For finite systems the linear optical absorption spectrum is obtained by applying a small momentum *k* to the electron, i.e. exciting all frequencies of the system, and then propagating these excited wave functions. The optical spectra are calculated through the dipole strength function:

$$S(w) = \frac{2w}{\pi} Im\left\{\frac{1}{k} \int dt\, e^{iwt}\, [d(t) - d(0)]\right\},$$

where the quantity inside the curled parenthesis is essentially the Fourier transform of the dipole moment of the system. The calculation of the dielectric function for a periodic system is performed in a similar way, considering the time-propagation of the Kohn-Sham orbitals after perturbing the system with a sudden change of the vector gauge Field A, following the scheme proposed by Bertsch *et al*.[69].

## 2.1 Assessment of the PBE method through comparison with range-separated hybrid functional and PCM model

The method of choice in this work is the PBE functional. However, it is generally known that it has some limitation, such as for example it underestimated the band gap value of semiconductors. In this section, in order to assess the reliability of the mechanism of charge transfer observed with PBE method, we present a comparative analysis with the more sophisticated range-separated hybrid functional CAM-B3LYP [59] and with the inclusion of the solvent effect through the PCM model [60]. However, CAM-B3LYP is computationally very costly, therefore we could not afford calculations on the same realistic 2.2 nm $TiO_2$ nanospheres (of 700 atoms) but we had to limit our comparative study to a small cluster of six $TiO_2$ units. We adsorbed one dopamine molecule on the cluster and performed conventional frequency domain or line response TDDFT calculations with the Gaussian16 suite [70] and 6-311++G* basis set. We first checked and confirmed that LR-TDDFT absorption spectrum is consistent with RT-TDDFT by Octopus code with the PBE functional (**Figure S2 and S3**). Then we compared the results of the two density functional methods (PBE vs CAM-B3LYP), with and without inclusion of the solvent effects (water), using the LR-TDDFT approach. All the results are reported in the Supplementary Material (**Tables S1 and S2**, **Figures S1-S6**), together with a detailed discussion and analysis.

A summary of the main results of this assessment study follows: 1) for isolated molecules, absorption spectrum with PBE does not differ very much from that with CAM-B3LYP and it is slightly in better agreement with experiments; 2) for isolated molecules, the inclusion of implicit water through the PCM model causes a tiny blue shift in the position of the bands with both PBE and CAM-B3LYP functionals; 3) for the isolated cluster, CAM-B3LYP absorption onset is shifted to higher energies with respect to the PBE one, as expected since it contains some exact exchange that better reproduces the band gap of the material; 4) for the isolated cluster, the inclusion of the solvent effect causes a blue-shift of the absorption spectra with both PBE and CAM-B3LYP functionals; 5) for the dopamine-$TiO_2$ cluster, we observe a similar shape of the differential absorption spectrum for CAM-B3LYP with respect to PBE, although it is shifted to higher energies, both in vacuum and in water.

Therefore, the conclusion we could draw from the assessment analysis is that the PBE method, even without inclusion of the solvent effect, is capable of catching the main physics of these complex systems, in particular the type of mechanism of charge transfer, since it provides a differential absorption spectrum for the dopamine-$TiO_2$ complex with similar features that one would obtain if using a more costly range-separated hybrid functional method and an implicit solvent model. The only difference is a rigid shift to lower energies that one should keep in mind when using the PBE method, as we will do in all this work.

## 3. RESULTS AND DISCUSSION

### 3.1 Optical properties of isolated dopamine and DOPAC molecules.

In this section, we will present the simulated absorption spectra of the two isolated linker molecules considered in this work, i.e. dopamine (black) and DOPAC (red), see **Figure 1.** These are obtained using a rather large broadening (short propagation time) in order to be comparable with experiments. However, in **Figure S7** in the Supplementary Material we also present a comparison with increasing propagation times. We can observe that the two spectra lines have a similar profile. The only difference between the two catechol-like molecules is the additional functional group (ethyl-amino for dopamine and carboxyl for DOPAC). In both cases, there are two main peaks of absorption: for dopamine the first one, less intense, is centred at 4.75 eV, while the second at 6.09 eV in good agreement with the experimental values of 4.43 eV and 5.63 eV, respectively [71]; for DOPAC the two main peaks are centred at 4.10 eV and 6.26 eV while in the experimental spectrum they are at 4.20 eV and 5.22 eV, respectively [72]. The onset of the absorption we extract from the first peak of the calculated spectra with the largest propagation time (5.4 fs, see **Figure S1**) is 4.22 eV for dopamine and 4.05 eV for DOPAC, which are a bit larger than the HOMO-LUMO gap values obtained at the same level of theory (PBE), of 4.06 eV and 3.81 eV, respectively.

In **Figure S8** in the Supplementary Material we show the separated contributions for the three different directions of propagation (x,y,z). For both molecules, the main contribution to the total spectrum comes from the x direction, which is the one in the ring plane passing exactly in between the hydroxyl groups on one side and going towards the amine (or carboxyl) functionality on the other.

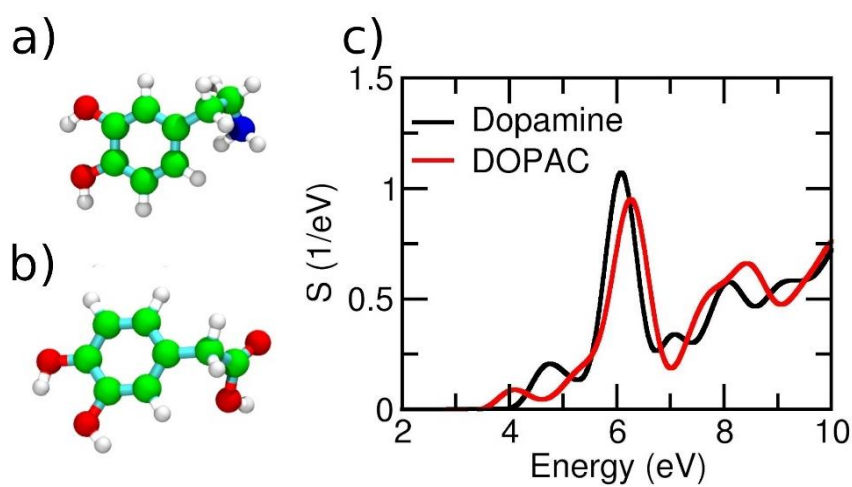

**Figure 1**: Ball-and-stick representation of the molecular structure of a) dopamine and of b) DOPAC. Red spheres represent O atoms, green spheres represent C atoms, blue sphere represents N atom and white small spheres represent H atoms. c) Calculated TDDFT absorption spectra for the isolated dopamine and DOPAC molecules. S is the strength function.

**3.2 Optical properties of the TiO$_2$ NP.**

The spherical NP used in this work with a diameter of 2.2 nm (233 TiO$_2$ units) has been obtained through a simulated annealing process followed by full atomic relaxation at the SCC-DFTB level of theory in a previous work by some of us [22,23]. Curved NPs of similar shape and size were used in several experimental studies for complexation with catechol-like ligands [28,47,49,51,53]. In **Figure 2a** we report the calculated absorption spectrum of this TiO$_2$ NP: it is continuous spectrum with a maximum intensity at 10 eV, in excellent agreement with previous reports [73,74]. For example, using the TDDFT method, Auvinen et al. reported the optical spectra of a series of (TiO$_2$)$_n$ clusters n = {1, 2, 8, 18, 28, 38}, which were all characterized by a maximum intensity peak centred at 10 eV, similarly to our spectrum, although other absorption details were found to depend on the shape and size of the cluster. In particular, they observed a smooth transition from a quite spiky spectrum for smallest cluster to a continuous spectrum for the largest one, i.e. a change from molecular-like electron structure, where occupied (valence) and unoccupied (conduction) bands are quite narrow and composed of almost molecular orbitals, towards the bulk-like electron structure with wide bands structure [73]. In another recent study [75], the simulated spectra from the frequency-dependent dielectric function for $(TiO_2)_n$ clusters of n = {1-20, 35, 84} were calculated showing discrete peaks for the smallest nanoclusters, whereas large continuum absorption bands for the largest ones.

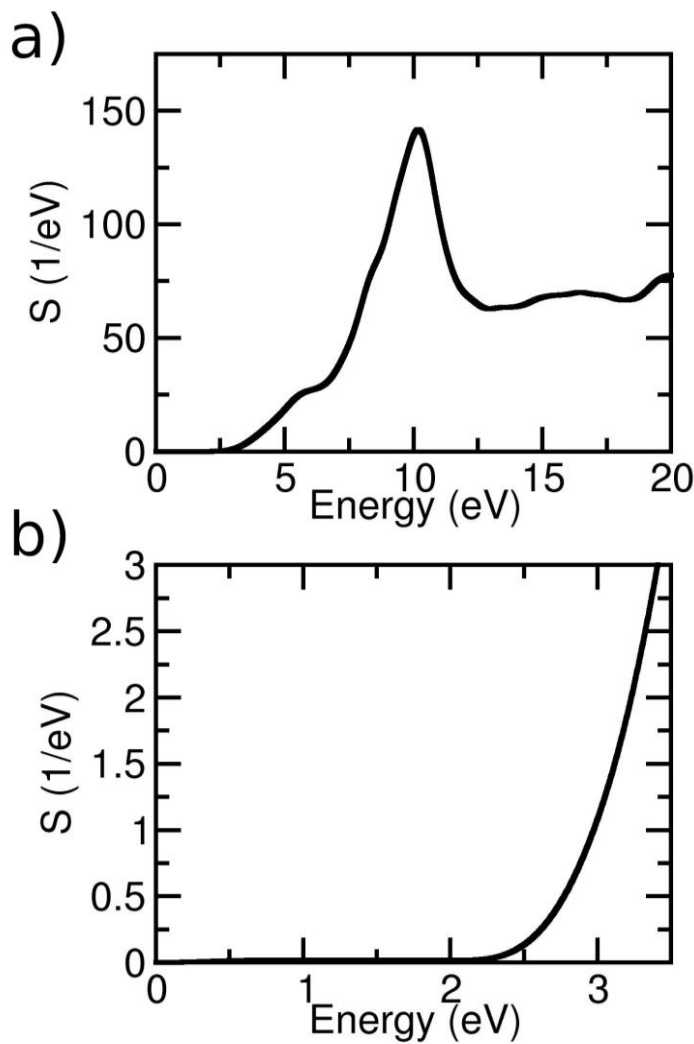

**Figure 2**: a) Absorption spectra from TDDFT calculations of the spherical $TiO_2$ NP with diameter of 2.2 nm. b) Zoom in of absorption spectrum in the region of the absorption edge.

In **Figure 2b** we present a closer inspection at the absorption edge. We extract from the calculated spectrum an $A_{onset}$ of 2.3 eV. This value is in agreement with previous reports of optical gap values for different $TiO_2$ clusters using the PBE functional [72,73]: for clusters from 1 to 20 units of $TiO_2$ the average of optical gap, using frequency-dependent dielectric function is 2.75±0.34 eV [75], whereas for bigger clusters, with more than 150 $TiO_2$ units, the Kohn−Sham gap was determined to be ~2.4 eV [76]. It is well-known that the standard gradient corrected functionals, such as PBE, produce an underestimated band gap value for transition metal oxides [75], which can be improved with the use of hybrid functionals or of many-body perturbation techniques, as in the GW and BSE methods [76-83]. However, the main goal of this work is not to achieve the most accurate reproduction of the band gap, but it is to unravel the nature of the charge transfer mechanism through the differences between the spectrum of the $TiO_2$-molecule complex and that of the isolated systems ($TiO_2$ NP on one side and cathecol-like molecule on the other).

### 3.3 Optical properties of modified $TiO_2$ NP.

In this section, we present the absorption spectra of the cathecol-like molecules anchored to the spherical anatase $TiO_2$ nanoparticle. First, we present two different adsorption configurations of dopamine (NP/DA), where it is either laying down on the surface or standing up towards the vacuum. Then, we present adsorbed DOPAC (NP/DC) in a standing up configuration.

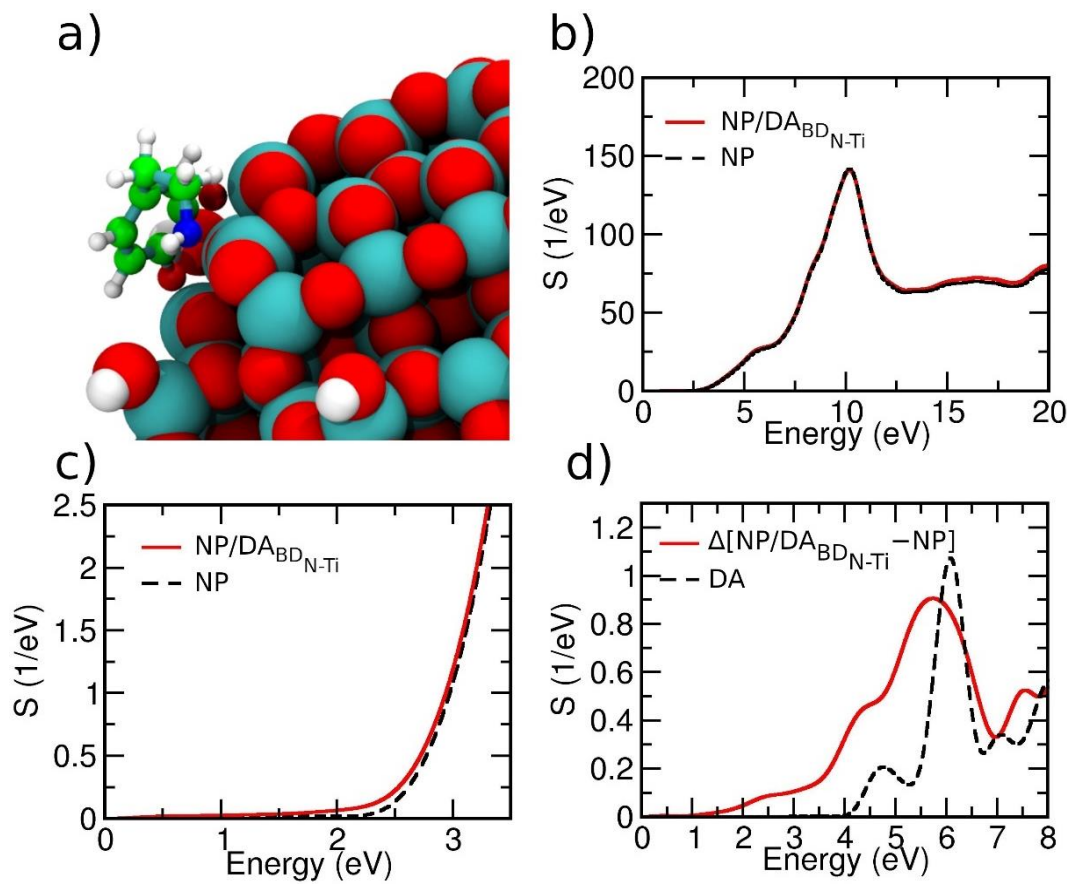

**Figure 3**: a) Ball-and-stick representation of the dopamine molecule adsorbed in BD$_{Ti\text{-bond}}$ configuration on the spherical NP in a space filling representation. b) Whole absorption spectrum of the NP/DA(BD$_{N\text{-Ti}}$) complex (red). For comparison, the spectrum of bare NP is also shown (dashed black line). c) Zoom in of absorption spectrum showed in b) in the region of the absorption edge. d) Differential absorbance spectrum ($\Delta$[NP/DA(BD$_{N\text{-Ti}}$)−NP]) in red. The absorption spectrum of a dopamine molecule is also shown for comparison (dashed black line).

### 3.3.1 Dopamine linker anchored to the TiO$_2$ NP in the BD$_{N\text{-Ti}}$ configuration

In a previous work we showed that, in the low coverage regime, dopamine can anchor the TiO$_2$ NP surface in many different ways [27]. Dopamine molecules favourably adsorb through dissociation of the two OH groups. The two dissociated protons are transferred to O$_{2c}$ atoms. The anchoring may occur either in a bidentate ("BD") or in a chelated ("C") fashion in agreement with experimental observations [42,84]. We found that most stable configuration is when the molecule is adsorbed in a bidentate fashion and it is bent toward the surface to establish a coordination bond between the amino N atom and a surface five-fold Ti atom (named BD$_{N\text{-Ti}}$) [27]. For the PBE-optimized structure of this configuration (**Figure 3a**) with a binding energy of −4.15 eV, we calculated the absorption spectrum by TDDFT method (**Figure 3b**). For comparative purposes, we also show in dashed line the absorption spectrum of the bare NP from the previous section. Looking at the shapes of the spectra it may seem that there is no difference between them. However, a closer inspection at the absorption edge region in **Figure 3c** allows appreciating a small red shift for the NP/DA suggesting a charge transfer mechanism.

In previous works, to prove a direct mechanism of charge transfer by means of TDDFT calculations, authors have highlighted the presence of a new band in the TiO$_2$-cluster/linker spectrum when compared with the spectral line of the isolated linker [34-36,39]. In our case, the use of a realistic NP of 233 units of TiO$_2$ makes this comparison not possible because the spectrum of NP/DA is continuous (i.e. there are no discrete peaks) and the absorption intensities of DA and NP/DA are very different. For this reason, to highlight the new absorption features in the NP/DA spectrum, we subtracted the NP spectrum from the NP/DA one (i.e. Δ[NP/DA−NP]). An analogous procedure has been previously used to analyse experimental spectra, for example, of common dyes adsorbed on silver nanoparticles [85]. The resulting spectral line is called "differential absorbance spectrum".

In **Figure 3d** we show the differential absorbance spectrum for the (BD$_{N-Ti}$) configuration (Δ[NP/DA(BD$_{N-Ti}$)−NP]). It is now possible to make a direct comparison with the spectrum of an isolated dopamine molecule and discuss how it has been modified by the presence of the NP underneath. First, we notice a broadening of the absorption peaks and a tiny red-shift in the position of characteristics peaks at 4.75 eV and 6.09 eV. However, the most striking new feature is the appearance of a new band between 1.5 eV and 3.5 eV. This is a clear evidence of a charge-transfer excitation from the dopamine into the NP conduction band in a direct injection mechanism (type-II).

Another interesting observation is that the main peak at ∼6 eV has a lower intensity than the corresponding peak of the isolated molecule. Dopamine molecule is characterized by a uniaxial transition dipole moment in the plane of π-conjugated benzene ring. Then, the observed reduced absorption can be attributed to the low-lying downward adsorption mode of dopamine (BD$_{N-Ti}$) with the π-ring of the dopamine almost parallel to the NP surface. This effect is similar to surface-selection rules in surface-enhanced spectroscopies [85-87] and has been also observed in the adsorption spectrum of the rhodamine 700 dye on silver nanoparticles [85].

Now we comment on the experimentally observed shift of the absorption onset from the UV to the visible region. At low dopamine coverage (5%), this shift is from 380 nm (bare NP) to 650 nm (NP/DA), i.e. of 1.36 eV. Increasing the coverage of dopamine to 20% or 50% on the NP enhances the absorbance at all wavelengths <800 nm but does not shift the absorption onset to a value larger than 800 nm. This translates into a shift between 1.36 eV and 1.6 eV of the A$_{onset}$ with respect to the bare NP [28]. In our calculations, having one dopamine molecule on the NP corresponds to a coverage of ∼2 %. The computed red-shift of the A$_{onset}$ for the (BD$_{N-Ti}$) configuration is of 0.8 eV (from 2.3 eV for the bare NP to 1.5 eV for NP/DA($BD_{N-Ti}$)), which is close to the experimental value of 1.36 eV at low coverage but slightly lower. For this reason, we have investigated another dopamine adsorption configuration on the NP that will be presented in the next section.

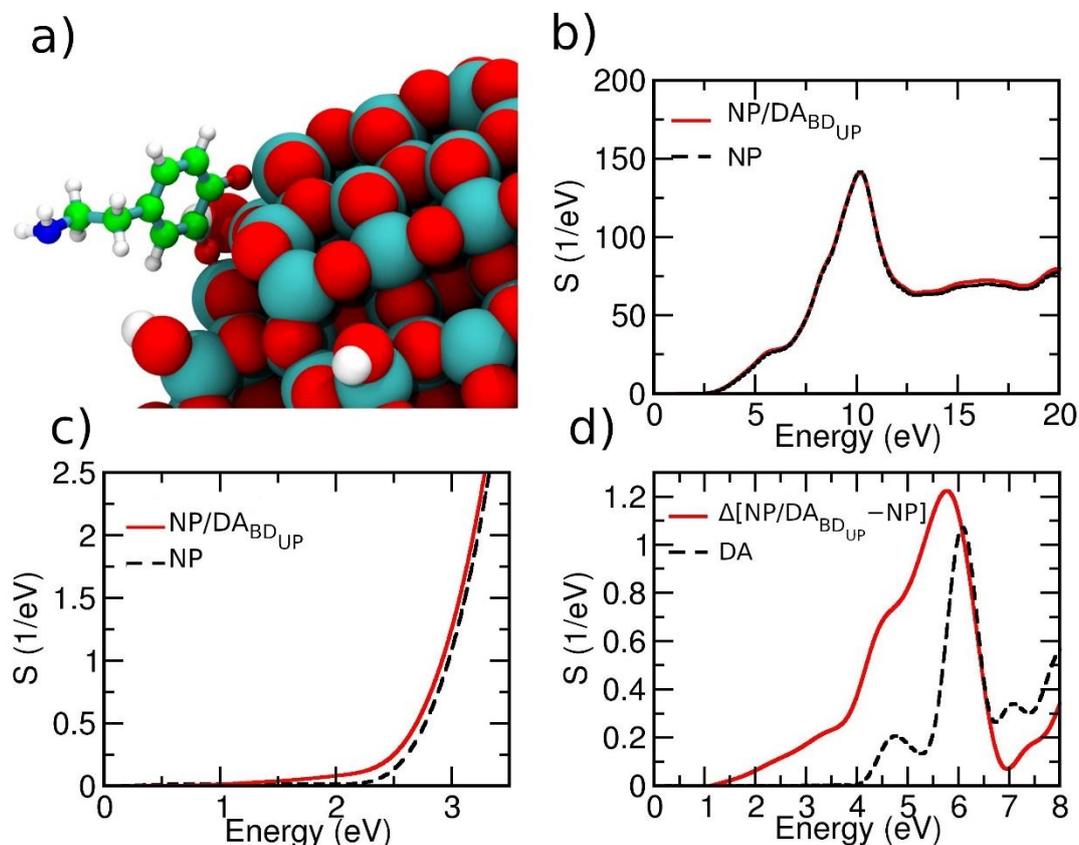

**Figure 4**: a) Ball-and-stick representation of the dopamine molecule adsorbed in $BD_{UP}$ configuration on the spherical NP in a space filling representation. b) Whole absorption spectrum of the NP/DA($BD_{UP}$) complex (red). For comparison the spectrum of bare NP is added (dashed black line). c) Zoom of absorption spectrum showed in b) in the region of absorption edge. d) differential absorbance spectrum (Δ[NP/DA($BD_{UP}$)−NP]) in red. The absorption spectrum of a dopamine molecule is added for comparison (dashed black line).

### 3.3.2 Dopamine linker anchored to the TiO₂ NP in the $BD_{UP}$ configuration

In this section we present an adsorption configuration (named $BD_{UP}$), where the dopamine molecule stands up, pointing the amino functional group toward the vacuum (**Figure 4a**). This would be the most common configuration of dopamine at large coverages [26,27] and probably also in an aqueous environment. **Figure 4b** shows the whole calculated absorption spectrum of this other configuration as compared with that of the bare NP. They look similar; however, a closer inspection of absorption edge region (in **Figure 4c**) shows a clear red shift of 1.0 eV, which is larger than that calculated for the downward configuration ($BD_{N-Ti}$), and is close to the experimentally observed shift of 1.36 eV, which was discussed at the end of the previous section. Some difference with respect to the experiment is due to a different coverage, where the minimum coverage is 5 %, which would be too computational expensive to simulate. As observed before for (BD$_{N-Ti}$) in **Figure 3d**, the differential absorbance spectra for (BD$_{UP}$) in **Figure 4d** shows not only a broadening of the absorption peaks and a tiny red-shift in the position of characteristics peaks of dopamine, with respect to the isolated molecule) but also a new feature between 1.25 eV and 3.5 eV. The last observation confirms also in this case a charge transfer in a direct injection mechanism (type-II). Differently from the (BD$_{N-Ti}$), however, the peak at 6 eV is enhanced in comparison with that for the isolated dopamine molecule. This effect is due to the different orientation of the π-ring of dopamine with respect to the NP surface. In the standing up adsorption mode of dopamine considered here, the transition dipole moment of the molecule is perpendicular to the surface, and thus, leads to an absorption enhancement compared with the isolated molecule [85].

### 3.3.3 Dopamine linker anchored to the TiO$_2$ NP in the BD$_{H-bond}$ configuration

In order to reinforce our claims, we also performed a TDDFT calculation of a bidentate dopamine bent toward the surface, which establishes a H-bond between the N atom of the NH$_2$ group and a surface OH, as is showed in **Figure 5a**. We named this configuration BD$_{H-bond}$. In this case, we observe a similar behaviour as for the BD$_{N-Ti}$, although with a smaller red shift (the calculated A$_{onset}$ is now ~1.7 eV vs 1.5 eV for BD$_{N-Ti}$). The differential absorbance spectrum (**Figure 5a**) also shows a broadening and a tiny red shift of the two characteristic dopamine peaks, together with a new feature between 1.7 eV and 3.5 eV. The peak at 6 eV in the differential absorbance spectrum has a lower intensity than the peak of the isolated dopamine molecule. These observations indicate a charge transfer in a direct injection mechanism (type-II) and a reduced absorption due to the low-lying downward adsorption configuration of dopamine with the π-ring almost parallel to the surface.

### 3.3.4 DOPAC linker anchored to the TiO$_2$ NP in the BD$_{UP}$ configuration

In this last section on the nanoparticle, we have considered the anchoring of a DOPAC molecule to the TiO$_2$ NP surface, in order to compare the effect of a different functional group on the catechol-like molecule. Since we did not want to compare difference due to a different second binding of the bifunctional linker to the surface, we chose to compare the adsorption configurations with the molecule that is standing up and directing the second functional group (in the present case a COOH group) towards the vacuum. In **Figure 5b** (top) we show the optimized structure such configuration (BD$_{UP}$). In the case of DOPAC, the absorption edge is characterized by a red shift of **1.3 eV** (A$_{onset}$

~ 1 eV) with respect to the bare NP, which is even larger than the red shift of **1.0 eV** calculated for dopamine in the same BD$_{UP}$ configuration (see **Figure 4d**).

A similar phenomenon was observed in a study of the adsorption of catechol-like molecules, including dopamine and DOPAC, on ceria NPs [88].

Regarding to the differential absorbance spectra, we can see a similar behaviour than for NP/DA in the BD$_{UP}$ configuration. A broadening of the main molecular absorption peak at about 6 eV, almost without shift, and an enhanced absorption if compared with the isolated molecule are observed. Also, the same new feature between 1 eV and 3.5 eV is present, which suggests, also in the case of DOPAC, a direct mechanism injection.

Thus, we have several evidences that when the molecules are absorbed in a low-lying downward configuration (independent of the type of bond established between the terminal functionality of the molecule and the surface) the differential spectra is characterized by a lower intensity of the main absorption peak, as compared with the isolated molecule, whereas when the molecules are adsorbed in a standing up configuration (for both dopamine and DOPAC) with the functionality pointing toward the vacuum, the differential spectra is characterized by an enhanced absorption peak, as compared with the isolated molecule. These observed effects are in line with the surface-selection rules in surface-enhanced spectroscopies [85-87].

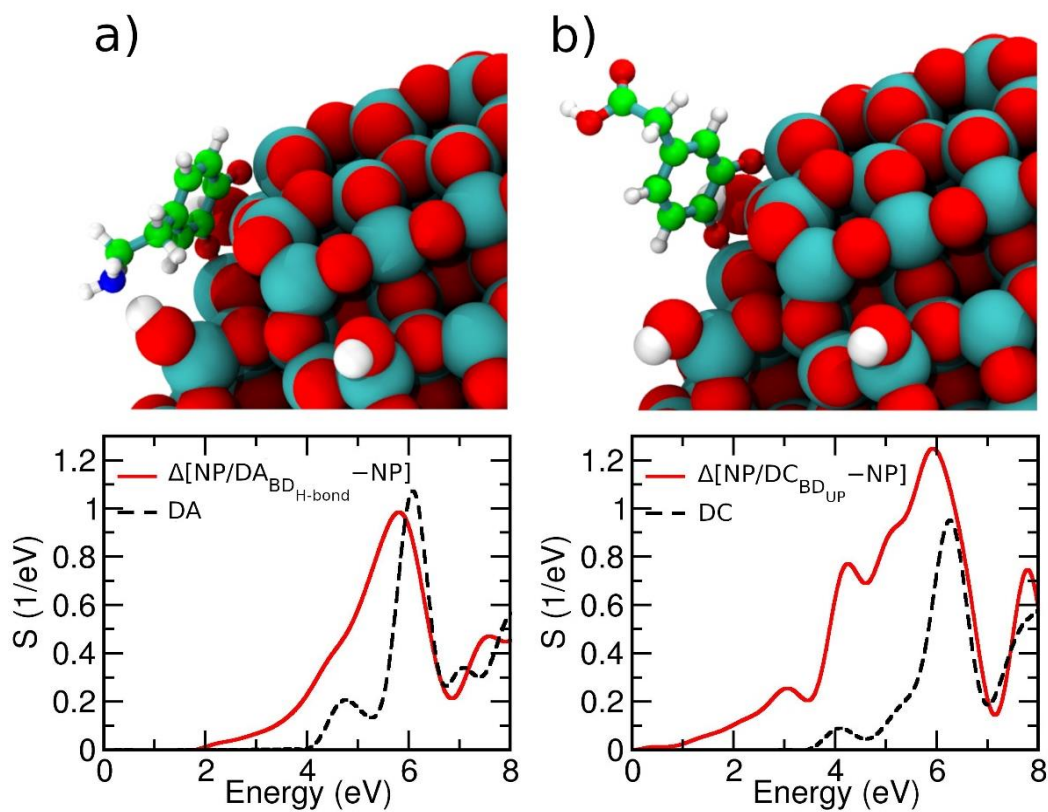

**Figure 5**: a) *Top-* Ball-and-stick representation of the dopamine molecule adsorbed in BD$_{H\text{-bond}}$ configuration on the spherical NP in a space filling representation. *Bottom-* Differential absorbance spectrum (Δ[NP/DA(BD$_{H\text{-bond}}$)−NP]) in red and dopamine molecule spectrum for comparison (dashed black line). b) *Top-* Ball-and-stick representation of the DOPAC molecule adsorbed in BD$_{UP}$ configuration on the spherical NP in a space filling representation. *Bottom-* differential absorbance spectrum (Δ[NP/DC(BD$_{UP}$)−NP]) in red and DOPAC molecule spectrum for comparison (dashed black line).

### 3.4 Optical properties of modified anatase TiO$_2$ (101) surface.

The existing TDDFT studies on modified TiO$_2$ surfaces were performed using TiO$_2$ clusters or small NPs [31-33,36,39-43]. Optical properties of catechol-like molecules bound to a flat TiO$_2$ surface have not been investigated so far due to technical difficulties in performing TDDFT on periodic systems. For this, in the present section we try to fill this gap and present a TDDFT study of the optical properties of modified (with dopamine and DOPAC) flat anatase (101) TiO$_2$ surface. As we mentioned in the **Computational Details Section** the simulated optical spectrum, in the case of periodic systems, is obtained by plotting the imaginary part of dielectric function vs. the photon energy. Given that for periodic system we have easy access to the electronic structure of the ground state through the density of states, we will also try to suggest which states of the molecule and of the semiconductor are probably involve in the lowest energy excitation in the simulated optical spectra.

#### 3.4.1 Dopamine adsorption at low coverage on anatase TiO$_2$ (101) surface

In order to simulate a sufficiently low coverage density of dopamine on the anatase (101) TiO$_2$ surface, we used a 2×4 supercell slab model, which results in a 0.25 ML coverage. We tested two configurations, in analogy to what has been done for the NP in the previous sections.

In the first configuration, S/DA(BD$_{N\text{-Ti}}$), the dopamine molecule bends toward the surface with the N atom of −NH$_2$ group coordinating to a surface five-fold Ti, as shown in **Figure 6a** (left panel). We determine the angle θ between the ring plane of the molecule and the surface plane to be only 20°. In the second configuration, or S/DA(BD$_{UP}$), the molecule is standing up with functional ethyl-amino group towards the vacuum, (**Figure 6a**, right panel). The angle θ in this case is measured to be 59°. We have given all the details about the variety of different possible adsorption configurations in a previous work by some of us in [63].

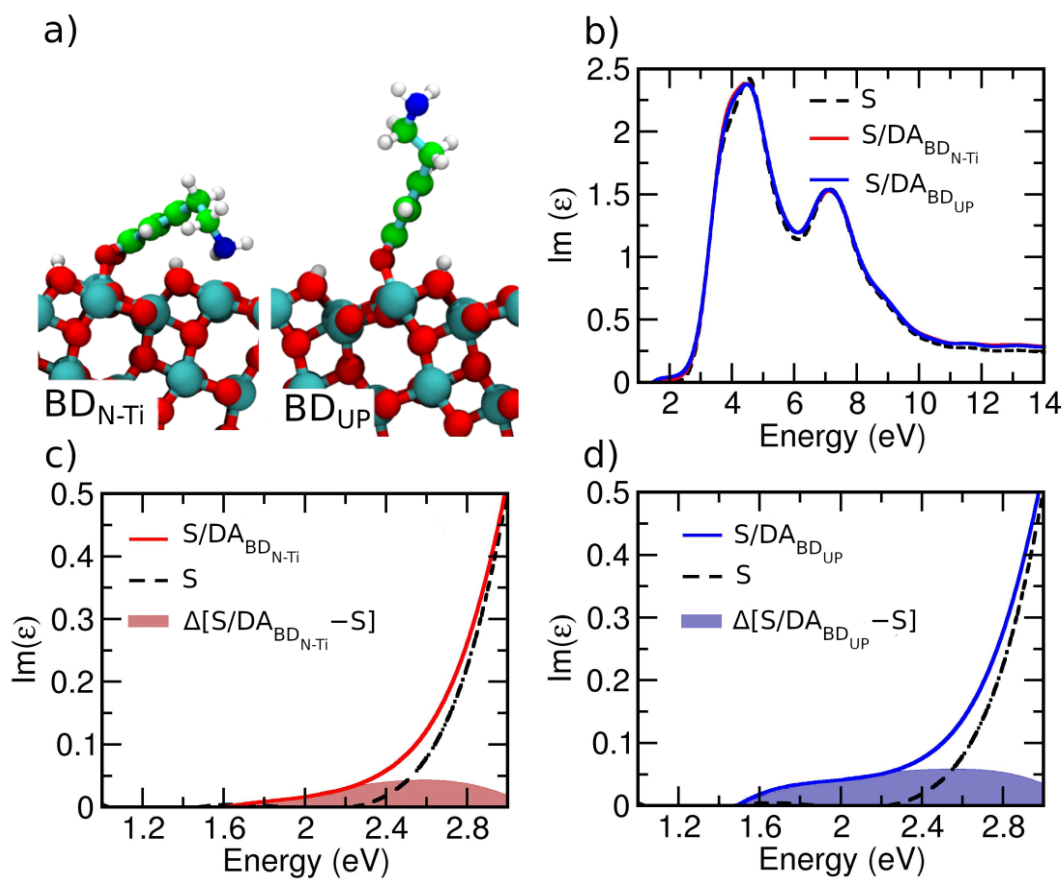

**Figure 6**: a) Ball-and-stick representation of the optimized geometries of adsorption of dopamine on the anatase (101) TiO$_2$ surface (S/DA) in low coverage regime. Left: BD$_{N-Ti}$ configuration. Right: BD$_{UP}$ configuration. b) Imaginary Part of the Dielectric Function of S/DA in the low coverage regime and of the bare TiO$_2$ (101) surface (S). Red BD$_{N-Ti}$, blue BD$_{UP}$, dashed black line bare surface. b) Zoom in the region of absorption edge for BD$_{N-Ti}$ configuration. Red area is differential spectrum ($\Delta$[S/DA (BD$_{N-Ti}$) –S]). c) Zoom in the region of absorption edge for BD$_{UP}$ configuration. Blue area is differential absorbance spectrum ($\Delta$[S/DA(BD$_{UP}$) −S]).

In **Figure 6b**, we present the simulated optical spectra of the bare (101) TiO$_2$ surface (dashed black line) and of the modified one (S/DA) in the two different configurations. The three spectra are very similar with two main peaks centred at 4.6 eV and 7.2 eV. However, a closer inspection of region around the absorption edge reveals a red shift of A$_{onset}$ after dopamine adsorption (see **Figures 6c and 6d**). The differential absorbance spectra (similar to what we did for NP/DA in the previous sections) present a broad feature starting from the lower absorption onset and covering a range of more than 1.5 eV. For BD$_{N-Ti}$ configuration it is 0.6 eV, while for BD$_{UP}$ configuration the shift is of 0.8 eV. **Figure 6c** shows that for BD$_{N-Ti}$ configuration the differential broad band starts at 1.70 eV, whereas for BD$_{UP}$ configuration at 1.50 eV (see **Figure 6d**). These results are in line with those obtained for functionalized spherical NPs in the previous sections: again, the red shift is larger for up configuration, which is due to the different inclination of the π-ring with respect to the surface, as already discussed above [85-87].

In order to get some further insight to rationalize the observed absorption peaks, we have also analyzed the electronic structure for the BD$_{UP}$ and the BD$_{N-Ti}$ configurations (see **Figure S9**). We have plotted the total (brown area) and projected density of states (DOS and PDOS) on TiO$_2$ (blue line), on the dopamine molecule (green area) and on the five-fold Ti atoms on the surface that bind with the anchoring DA O atoms (violet area) or with the N atom of the DA amino group (yellow area). For the bare TiO$_2$ surface, the HOMO-CB gap corresponds to the band gap (E$_{gap}$), calculated between the valence band maximum (VBM) and the conduction band minimum (CBM). For the S-DA systems, the HOMO is a mid-gap molecular state involving the C atoms of the π system, whereas the first empty state is the CBM. According to our present study, the charge transfer between the molecule and TiO$_2$ takes place through a direct injection mechanism (type-II), which we suppose is due to an excitation from the HOMO of the molecule and to some states at the bottom of the CB. We may notice that HOMO-CBM gap in the DOS and the absorption onset from the TDDFT calculations are different (**Table S3** in Supporting Information). We explain this by considering that the electronic transition is not to the CBM states but to some unoccupied Ti *d* states slightly higher in the CB, which are mixed with the dopamine π-states, i.e. the *d* states of the Ti atoms directly bonded to dopamine. To prove this and to gain more insight into the chemistry of the dopamine anchoring the TiO$_2$ (101) surface, we have analyzed the charge density difference, Δρ(r), defined as the local variation in the electron density with respect to the separated fragments (TiO$_2$ surface and dopamine), in the same atomic positions as when in the complex. In the Δρ(r) plots, shown in **Figure S9**, the red color represents electron charge accumulation and the blue color represents electron depletion regions. Indeed, we observe a depletion in the π orbital of dopamine and a net increase in density in the *d* states of the Ti atoms bound to dopamine. It is reasonable to expect that these states will play a major role in the electronic transition because there will an efficient coupling between the TiO$_2$ and DA states, improving the transition dipole moment and, thus, the probability of the transition, as discussed in the introduction. The difference between the energy of the *d* states of Ti atoms bound to the anchoring DA O atoms inside the CB and of the HOMO of DA (Δ[Ti$_{bound}$-HOMO]) for the BD$_{UP}$ and the BD$_{N-Ti}$ configurations are reported in **Table S3**: 1.56 eV and 1.89 eV, respectively. These values are consistent with the shift between the corresponding A$_{onset}$ values for the two systems (1.50 eV and 1.70 eV).

### 3.4.2 Dopamine adsorption at full coverage on anatase TiO$_2$ (101) surface

In a previous study by some of us, we investigated the growth of a full ML of dopamine molecules on the anatase (101) surface and we found out that, under kinetic conditions, two configurations are present on the surface [61] (**Figure 7a**) with a 50:50 proportion: the ML-BD$_{DOWN+H}$ (**Figure 7a**, left panel), in which all the DA molecules bend down and the NH$_2$ groups acquire one proton from the surface OH, and the ML-BD$_{UP}$ (**Figure 7a**, right panel), in which the ethyl-amino functional groups of all the DA molecules stand up towards the vacuum.

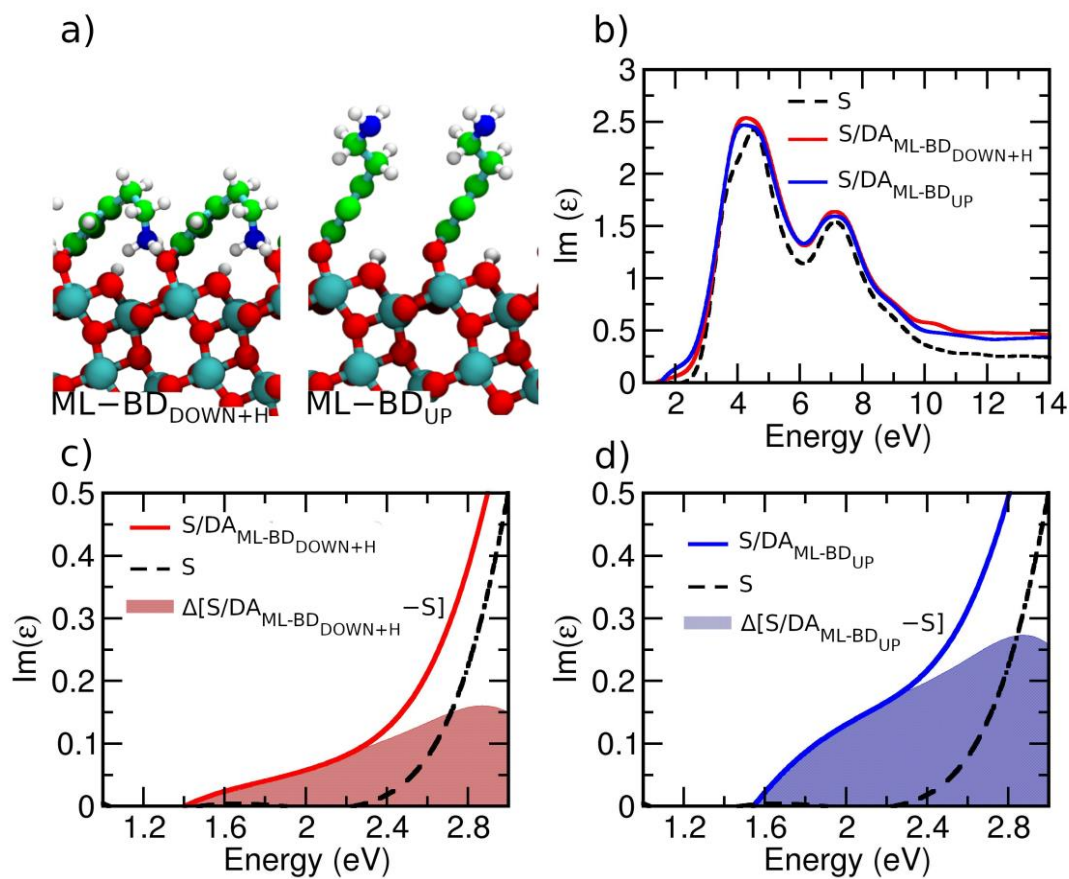

Figure 7: a) Ball-and-stick representation of the optimized geometries of adsorption of dopamine on the anatase (101) TiO$_2$ surface (S/DA) in full coverage regime. Left: ML-BD$_{DOWN+H}$ configuration Right: ML-BD$_{UP}$ configuration. b) Imaginary Part of the Dielectric Function of S/DA in the full coverage regime and of the bare TiO$_2$ (101) surface (S). Red ML-BD$_{DOWN+H}$, blue ML-BD$_{UP}$, dashed black line bare surface. c) Zoom in the region of absorption edge for ML-BD$_{DOWN+H}$ configuration. Red area is differential spectrum ($\Delta$[S/DA(ML-BD$_{DOWN+H}$)−S]). d) Zoom in the region of absorption edge for ML-BD$_{UP}$ configuration. Blue area is differential absorbance spectrum ($\Delta$[S/DA(ML-BD$_{UP}$)−S]).

In **Figure 7b** we present the spectra of both ML configurations and the spectrum of bare surface for comparison. With respect to the low coverage regime (see **Figure 6b**), we observe an increase in absorption intensity in the whole spectra and an increase in the intensity of the differential absorbance spectra as a consequence of the higher density of adsorbed molecules. However, analysing the absorption edge region (**Figures 7c** and **7d**), we notice that the A$_{onset}$ values for BD$_{UP}$ and ML-BD$_{UP}$ are very similar (1.50 eV vs. 1.54 eV), while the difference for BD$_{N-Ti}$ and for ML-BD$_{DOWN+H}$ is larger (1.70 eV vs. 1.42 eV). This is because in the first case, the low and full coverage standing up geometries are very similar (see right panels in **Figures 6a** and **7a**), with an angle between the π-ring of the molecule and the surface of 59º and 55º, respectively. On the contrary, when comparing the structural details of BD$_{N-Ti}$ and BD$_{DOWN+H}$ we may notice a larger difference in the angle θ between the molecular π-ring plane and the surface plane: 28.5º vs 45º, respectively. Therefore, we conclude that the geometry of adsorption of the linker on the surface plays an important role in the resulting optical properties of the modified TiO$_2$. As we mentioned above, for BD$_{UP}$ and ML-BD$_{UP}$ the main difference is in the absorption intensity, whereas the A$_{onset}$ values are very close. This is in agreement with the experimental spectra for NP/DA, which show that when the coverage of DA on the NP increases from 20% to 50% or more, the red shift of the absorption onset (with respect to bare NP) is rather constant, whereas the absorbance is enhanced [28].

### 3.4.3 DOPAC at low and full coverage on anatase TiO$_2$ (101) surface

In this section, we present the study of the optical properties for DOPAC-functionalized anatase (101) TiO$_2$ surface, with the main purpose to obtain to compare them with what reported above for dopamine. We considered one adsorption configuration (BD$_{UP}$) at both low and full coverage, as shown in **Figure 8a**. The π-ring plane of the molecule forms an angle with the surface of 65º and 61.5º for low and full coverage, respectively. In **Figure 8b** we report the whole spectrum for the two coverage regimes. The shapes of the curves are similar to that for the bare surface. However, a larger intensity in the absorption is observed for full coverage. A closer inspection of the absorption edge region (**Figure 8c** and **8d**) shows the same A$_{onset}$ of 1.40 eV for both coverages, which is expected because the relative orientation of the molecule with respect to the surface is very similar at the two coverages. This value is slightly red shifted if compared to the adsorption onset of dopamine in the same upward configuration (1.50 and 1.54 eV at low and full coverage, respectively), similarly to what observed for the molecules when adsorbed on TiO$_2$ NP in the previous sections. We attribute these differences in the shift to the fact that the free DOPAC molecule itself absorbs at lower energies than dopamine (see Figure 1). Regarding to the differential absorbance spectra also reported in **Figure 8c** and **8d**, we may notice that an increase in molecular density results in an increase in the absorption intensity but does not cause a relevant absorption shift, which is also observed for dopamine in the same upward configuration (1.50 and 1.54 eV at low and full coverage, respectively).

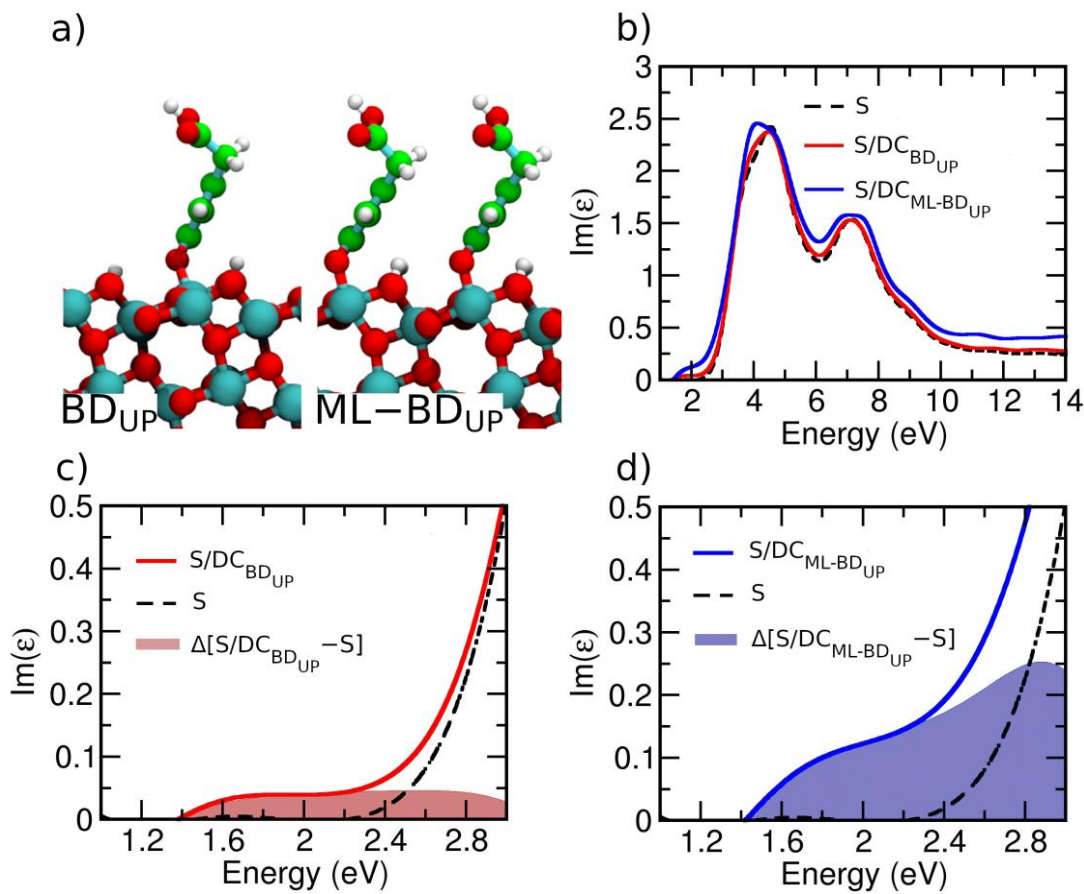

**Figure 8**: Ball-and-stick representation of the optimized geometries of adsorption of DOPAC on the anatase (101) $TiO_2$ surface (S/DC) in low ($BD_{UP}$), (left panel), and full (ML-$BD_{UP}$) (right panel) coverage regime. b) Imaginary Part of the Dielectric Function of S/DC in the low (red) and full (blue) coverage regime and of the bare $TiO_2$ (101) surface (S, dashed black line). c) Zoom in the region of absorption edge for $BD_{UP}$. Red area is differential spectrum ($\Delta[S/DC(BD_{UP})-S]$). d) Zoom in the region of absorption edge for ML-$BD_{UP}$. Blue area is differential absorbance spectrum ($\Delta[S/DC(ML-BD_{UP})-S]$).

## 4-CONCLUSIONS

In this work, we have performed an extensive TDDFT investigation of the optical properties of catechol-like molecules adsorbed on curved and flat anatase $TiO_2$ surfaces. We used a realistic $TiO_2$ NP with diameter of 2.2 nm and a periodic slab model, on which we have anchored dopamine and DOPAC at different coverage. The method of choice was the PBE functional, whose reliability has been assessed through comparison of results, for a dopamine molecule adsorbed on a small cluster of six $TiO_2$ units, with those obtained using a more sophisticated range separated hybrid functional (CAM-B3LYP) and by adding the solvent effect through the PCM model.

For both dopamine- and DOPAC-modified NPs, we have observed a red shift (DOPAC > dopamine) in the absorption spectrum if compared with the one for bare NPs, in line with the experimental observations. [28,29] The differential absorbance spectrum, where the contribution from the NP is removed, shows clearly an additional new band that is not present in the spectrum of the isolated molecule. This new feature is at lower energies than the molecular excitation of the free molecule, which is an evidence of a direct charge transfer or type II mechanism of injection.

In addition, we found a relation between the configuration of molecular adsorption and the intensity of the main peak in the differential absorbance spectra. When the molecule is adsorbed in a laying down configuration (with the π-ring almost parallel to the surface), the intensity of the main peak is lower than for the isolated molecule, whereas in a standing up configuration (with the π-ring almost perpendicular to the surface), it is larger. This effect can be rationalized applying the same surface-selection rules of the surface-enhanced spectroscopies [82-84].

As regards the dopamine- and DOPAC-modified flat $TiO_2$ surfaces, we carried out, for the first time, a TDDFT investigation of the periodic slab models with the real-time and real-space method. Similarly, to what done for NPs, we could present the absorption spectra, differential absorbance spectra but, additionally, we could also calculate the projected density of states, which are found to be useful to interpret the direct charge transfer injection mechanism. In the case of flat surfaces, we could also investigate the molecular coverage effect going from 0.25 to 1 ML.

In general, we found that the larger red shifts are produced when the linker is absorbed in standing up configurations, where the angle between the π-ring of the molecule and the surface plane is large than 45º. Regarding to the coverage the main difference in related with a higher intensity of the whole spectra for full coverage respect to the low coverage.

The excitations that are responsible for the absorption onset red shift are characterized by lower energy values than any of the free molecule excitations. Therefore, we supposed that these low energy excitations in the spectrum of dopamine- or DOPAC-modified $TiO_2$ complexes are due to electronic transitions from the molecular HOMO to some empty states in the CB. In order to have the highest intensity of absorption, and thus, the largest transition dipole moment, we expect that the empty states involved are not those at the very bottom edge of the CB, but the d states of the Ti atoms, which are mostly coupled with the molecular states of the linker. According to the PDOS analysis, those states are somewhat higher in the conductions band. The one-electron energy differences we can extrapolate from the PDOS agree rather well with the absorption onset values both for the low and for the full coverage regimes.

**Acknowledgments**

The authors are grateful to Arrigo Calzolari for fruitful discussions. The project has received funding from the European Research Council (ERC) under the European Union's HORIZON2020 research and innovation programme (ERC Grant Agreement No [647020]).

**Supplementary Material**

Supplementary data associated with this article can be found in the online version at….

# Supplementary Material

**Absorption mechanism of dopamine/DOPAC modified TiO$_2$ nanoparticles**

**by time-dependent density functional theory calculations**

Costanza Ronchi,[1,2,†] Federico Soria,[1,†] Lorenzo Ferraro,[1] Silvana Botti,[2] Cristiana Di Valentin[1,*]

[1]Dipartimento di Scienza dei Materiali, Università di Milano Bicocca

Via R. Cozzi 55, 20125 Milano Italy

[2]Friedrich Schiller University Jena, Institut für Festkörpertheorie und -optik,

Max-Wien-Platz 107743 Jena, Germany

---

[†] These authors have equally contributed to this work

[*] Corresponding author: cristiana.divalentin@unimib.it

# Assessment of the PBE method through comparison with range-separated hybrid functional and PCM model

Based on the results in Table 1, we may conclude that PBE better reproduces the lowest band energy than CAM-B3LYP functional for both molecules. The inclusion of solvent effects through a PCM model modifies the excitation energies by about 0.1-0.2 eV only in the case of PBE calculations. Based on the results of Table 2 regarding the second absorption band, may conclude that both functionals tend to overestimated the energy of this excitation, however, PBE less than CAM-B3LYP. One may note some differences between PBE/Octopus and PBE/Gaussian16, which are due to the different setups (e.g. LR-TDDFT versus RT-TDDFT, different basis sets, all-electrons versus pseudopotentials, etc.).

**Table S1:** First lowest excitation energies (eV) for free Dopamine and DOPAC molecules from LR-TDDFT calculations using the PBE and CAM-B3LYP functionals, with and without the inclusion of the solvent effect through a PCM Model. The values from RT-TDDFT calculations are also shown.

|  | PBE | PBE(PCM) | CAM-B3LYP | CAM-B3LYP (PCM) | PBE Octopus | EXP |
|---|---|---|---|---|---|---|
| Dopamine | 4.58 (+0.15) | 4.81 (+0.38) | 5.01 (+0.58) | 5.01 (+0.58) | 4.75 (+0.32) | 4.43 |
| DOPAC | 3.94 (-0.26) | 3.88 (-0.32) | 5.01 (+0.81) | 5.01 (+0.81) | 4.10 (-0.10) | 4.20 |

**Table S2:** Second lowest excitation energies (eV) for free Dopamine and DOPAC molecules from LR-TDDFT calculations using the PBE and CAM-B3LYP functionals, with and without the inclusion of the solvent effect through a PCM Model. The values from RT-TDDFT calculations are also shown.

|  | PBE | PBE(PCM) | CAM-B3LYP | CAM-B3LYP (PCM) | PBE Octopus | EXP |
|---|---|---|---|---|---|---|
| Dopamine | 5.96 (+0.33) | 6.01 (+0.38) | 6.35 (+0.72) | 6.52 (+0.89) | 6.09 (+0.46) | 5.63 |
| DOPAC | 5.95 (+0.72) | 5.99 (+0.75) | 6.41 (+1.19) | 6.51 (+1.29) | 6.26 (1.04) | 5.22 |

In summary, we observed that for isolated molecules the choice of range-separated functionals does not improve considerably the position of the two main lowest bands.

In the following, we will compare and discuss the behavior of PBE vs CAM-B3LYP for a dopamine-TiO$_2$ complex. Before we show our results, we wish to recall that according to previous studies the use of PBE functional is a good compromise in terms of accuracy/computational cost for the description of the charge transfer processes in dye-TiO$_2$ systems [1,2]. CAM-B3LYP calculations are much more expensive. For this reason, we decided not to use our realistic TiO$_2$ NP of 700 atoms, which is prohibitive with CAM-B3LYP, but on a smaller cluster of 6TiO$_2$ units (**Figure S1**), as already proposed in the literature [1,3,4].

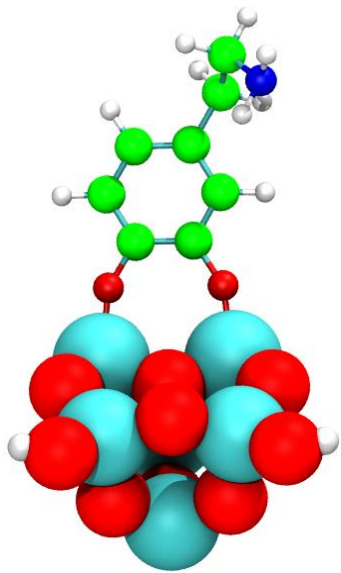

**Figure S1**: Ball-and-stick representation of the optimized dopamine molecule adsorbed on the 6(TiO$_2$) cluster.

First, we have confirmed that the charge transfer mechanism, which we found for the 2.2nm NS using PBE functional (RT-TDDFT with Octopus) is also observed when using the small TiO$_2$ cluster model. In **Figure S4a** we show the optical absorption spectrum for the 6(TiO$_2$) cluster (dashed black line) and for the 6(TiO$_2$)/DA complex (red line). Comparing with the 2.2 nm NP systems, the spectra show more discrete peaks for the 6(TiO$_2$) and also for the 6(TiO$_2$)/DA complex, which is expected due to the molecular-like electron structure where occupied (valence) and unoccupied (conduction) bands are quite narrow and composed of almost molecular orbitals. The shape of the spectra is in agreement with previous reports [5,6]. Even though, the shape of the TiO$_2$ spectrum change respect to the 2.2 nm NP we observed that the adsorption onset is 2.3 eV as we reported for the 2.2 nm NP in the manuscript. Upon dopamine adsorption on the TiO$_2$ cluster, a new onset in the absorption is observed at 0.6 eV, indicating a red shift in the absorption edge as for the 2.2 nm NP. **Figure S2b** shows the differential (i.e. $\Delta$[6(TiO$_2$)/DA$-$6(TiO$_2$)]) and dopamine spectra. We notice a broadening of the absorption peaks and a red-shift in the position of the peak at 4.75 eV. However, the most striking new feature is the appearance of new bands between 1.5 eV and 3.5 eV. This is the same behaviour that we have observed for dopamine anchored to the 2.2nm NP. Based on these results, we confirm that the absorption mechanism we observe for the realistic large system is also observed for the small cluster.

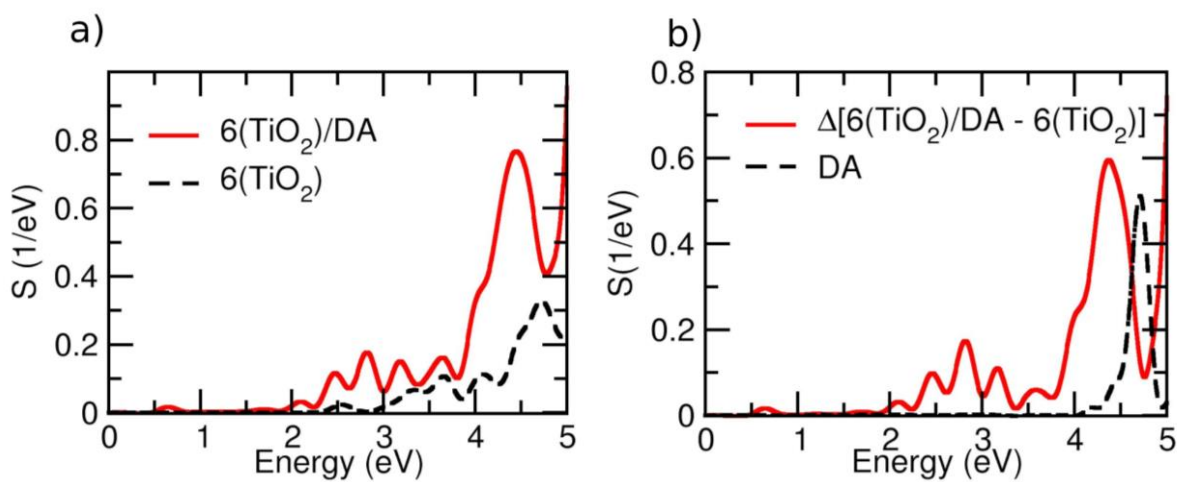

**Figure S2** a) Absorption spectrum obtained by RT-TDDFT (Octopus code) with the PBE functional for the 6(TiO$_2$)/DA complex (red line). For comparison, the spectrum of the isolated 6(TiO$_2$) cluster is also shown (dashed black line). b) Differential absorbance spectrum ($\Delta$[6(TiO$_2$)/DA−6(TiO$_2$)]) in red. The absorption spectrum of a dopamine molecule is also shown for comparison (dashed black line).

As a next step, we have repeated calculation of the spectra performing conventional LR-TDDFT calculations with Gaussian16 program. **Figures S3a and S3b** show the spectra of the 6(TiO$_2$) and 6(TiO$_2$)/DA complex and the comparison between the differential spectra and DA spectrum, respectively. It can see a very similar shape of the spectra in comparison with those obtained by RT-TDDFT (**Figure S2a**). The absorption onset for the cluster is 2.5 eV while for the complex the value is 0.6 eV. The differential spectra show a new band between 1.5 and 3.5 eV. The values indicate a very good agreement between the two TDDFT approaches: LR-TDDFT and RT-TDDFT.

One of advantages of LR-TDDFT is the possibility to characterize the nature of the states involved in the electronic transitions, as shown in **Figure S4** for the first absorption peak. The first band in the absorption spectrum of the TiO$_2$-dopamine complex at 0.6 eV is a HOMO→LUMO transition. The second band at 1.94 eV corresponds to a HOMO→LUMO+4 transition, whereas the peak at 2.27 eV is a HOMO→LUMO+7 transition and, finally, the peak at 2.64 eV is a HOMO-2→LUMO+4 transition. Clearly, all the excitations have a strong charge-transfer character, which is typically observed in the case of a direct or type II mechanism.

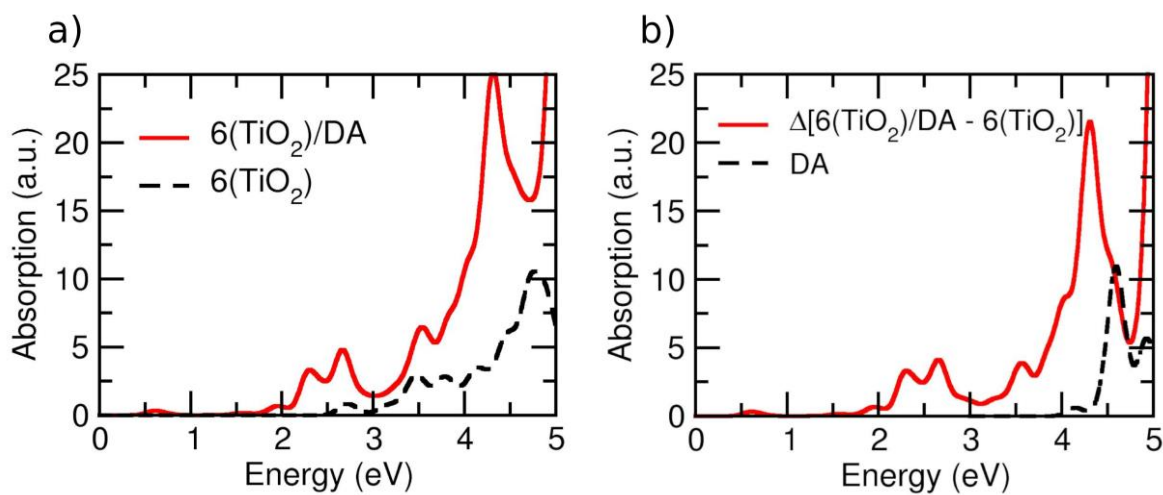

**Figure S3:** a) Absorption spectrum obtained by LR-TDDFT (Gaussian16 code) with the PBE functional for the $6(TiO_2)$/DA complex (red line). For comparison, the spectrum of the isolated $6(TiO_2)$ cluster is also shown (dashed black line). b) Differential absorbance spectrum ($\Delta[6(TiO_2)/DA-6(TiO_2)]$) in red. The absorption spectrum of a dopamine molecule is also shown for comparison (dashed black line).

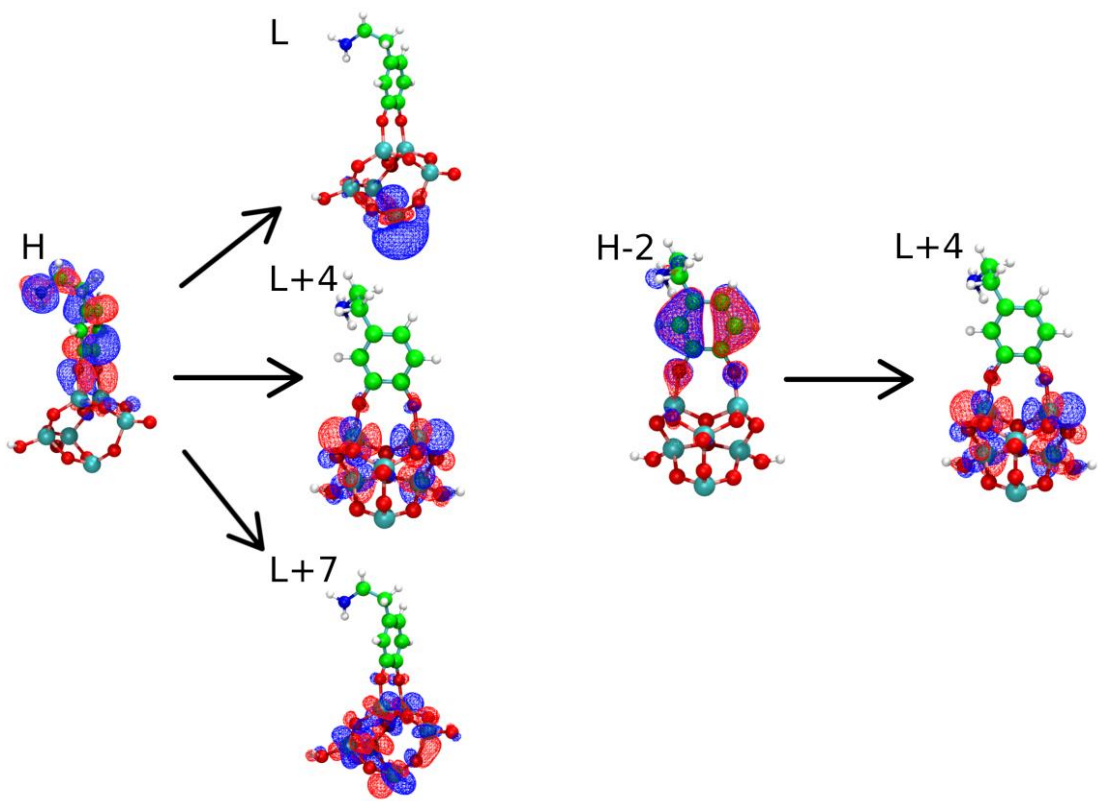

**Figure S4:** Selected transitions from occupied and to virtual molecular orbitals of low energy bands in the absorption spectrum, calculated with the PBE functional (Gaussian16) for dopamine bonded to the $6(TiO_2)$ cluster.

Now, we will show that the same absorption mechanism observed with PBE hold for CAM-B3LYP calculations. For this comparison we used LR-TDDFT calculations since they allow to compare both the spectra and the electronic states involved in the excitations.

In **Figure S5a** reports the optical spectra of the 6(TiO$_2$) cluster (dashed black line) and of the 6(TiO$_2$)/DA complex (red line). Clearly a blue shift in both cases is observed respect to the PBE calculation (**Figures S2a and S3a**). The absorption onset for the 6(TiO$_2$) cluster is 4.6 eV in a very good agreement with previous calculations [4]. When the dopamine is adsorbed on the cluster the adsorption onset moves to 3.5 eV, according to a red shift similar to what observe using the PBE functional. Analyzing the differential absorption spectrum and comparing it with that of isolated dopamine (**Figure S5b**), we notice a new band between 3.5 eV and 4.75 eV, which, again, is a clear evidence of a charge-transfer excitation from the dopamine into the NP conduction band states through a direct injection mechanism (type-II).

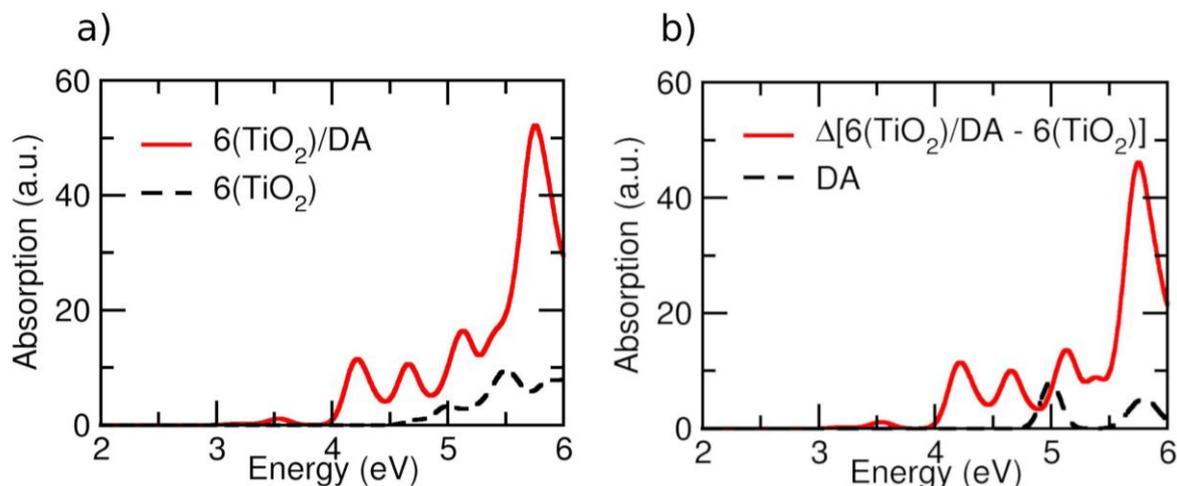

**Figure S5:** a) Absorption spectrum obtained by LR-TDDFT (Gaussian16 code) with CAM-B3LYP functional for the 6(TiO$_2$)/DA complex (red line). For comparison, the spectrum of the isolated 6(TiO$_2$) cluster is also shown (dashed black line). b) Differential absorbance spectrum ($\Delta$[6(TiO$_2$)/DA−6(TiO$_2$)]) in red. The absorption spectrum of a dopamine molecule is also shown for comparison (dashed black line).

To summarize and conclude, we have confirmed that calculations with CAM-B3LYP functional indicate the same mechanism of charge transfer that resulted from PBE calculations. We also observe a higher absorption onset in the spectra of the isolated cluster and of the complex when we move from PBE to CAM-B3LYP functional, a clear the red shift in the A$_{onset}$ of the complex respect to the isolated cluster is observed independently of the functional.

In order to investigate the effect of the solvent on the absorption spectra, we have performed additional calculations with the PCM model. For this comparative study, we used a smaller $TiO_2$ model. We compare results with and without solvent environment for both PBE and CAM-B3LYP functionals.

In **Figure S6** we report the spectra of the $6(TiO_2)$ cluster and the $6(TiO_2)$/DA complex using the PBE and CAM-B3LYP functionals with the PCM model (**Figures S8c and S8d**). For comparison we add the spectra with both functionals without PCM model (**Figures S6a and S6b**). A blue shift of the absorption onset of both functionals can be observed when the solvent effects are described. With the PBE functional, the $A_{onset}$ of the cluster moves from 2.3 to 3.1 eV and for the $6(TiO_2)$/DA complex the $A_{onset}$ moves from 0.6 to 1.5 eV. However, we can observe in **Figure S6c** that the charge transfer mechanism is not affected by the presence of the solvent since a clearly red-shift of adsorption spectrum is observed when the dopamine molecule is adsorbed in the $TiO_2$ cluster, which leads to a differential spectrum that is very similar to that in vacuum, except of a rigid shift. Using CAM-B3LYP functional we observe a behavior similar (compare **Figures S6b and S6d**). There is a blue-shift of the absorption onset for the cluster (from 4.6 to 4.8 eV) and for the complex (from 3.5 to 3.75 eV).

**Figures S6e and S6f** show the differential spectra compared with dopamine including the solvent effect. Also, it can see a new band a lower energies (between 1.7 and 4 eV for PBE and between 3.75 and 4.75 eV for CAM-B3LYP functional) showing a direct charge transfer mechanism.

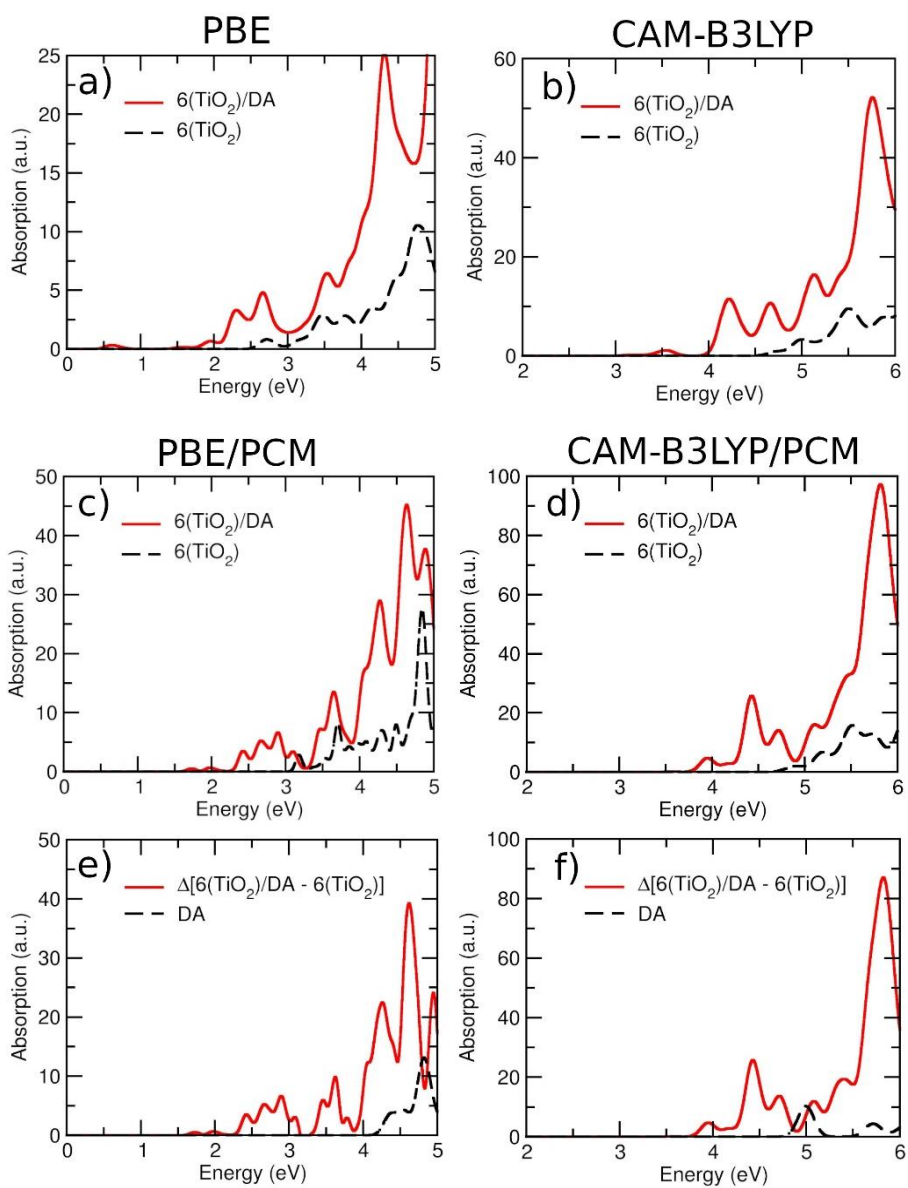

**Figure S6:** a) and b): PBE and CAM-B3LYP absorption spectra obtained by LR-TDDFT (Gaussian16 code) for the 6($TiO_2$)/DA complex (red line) and for the isolated 6($TiO_2$) cluster (dashed black line). c) and d) PBE and CAM-B3LYP absorption spectra obtained by LR-TDDFT (Gaussian16 code), including the solvent effect for the 6($TiO_2$)/DA complex (red line) and for the isolated 6($TiO_2$) cluster (dashed black line). e) and f) PBE and CAM-B3LYP differential absorbance spectra ($\Delta$[6($TiO_2$)/DA−6($TiO_2$)]) in red. The absorption spectrum of a dopamine molecule is also shown for comparison (dashed black line).

**Optical spectra of dopamine and DOPAC molecules**

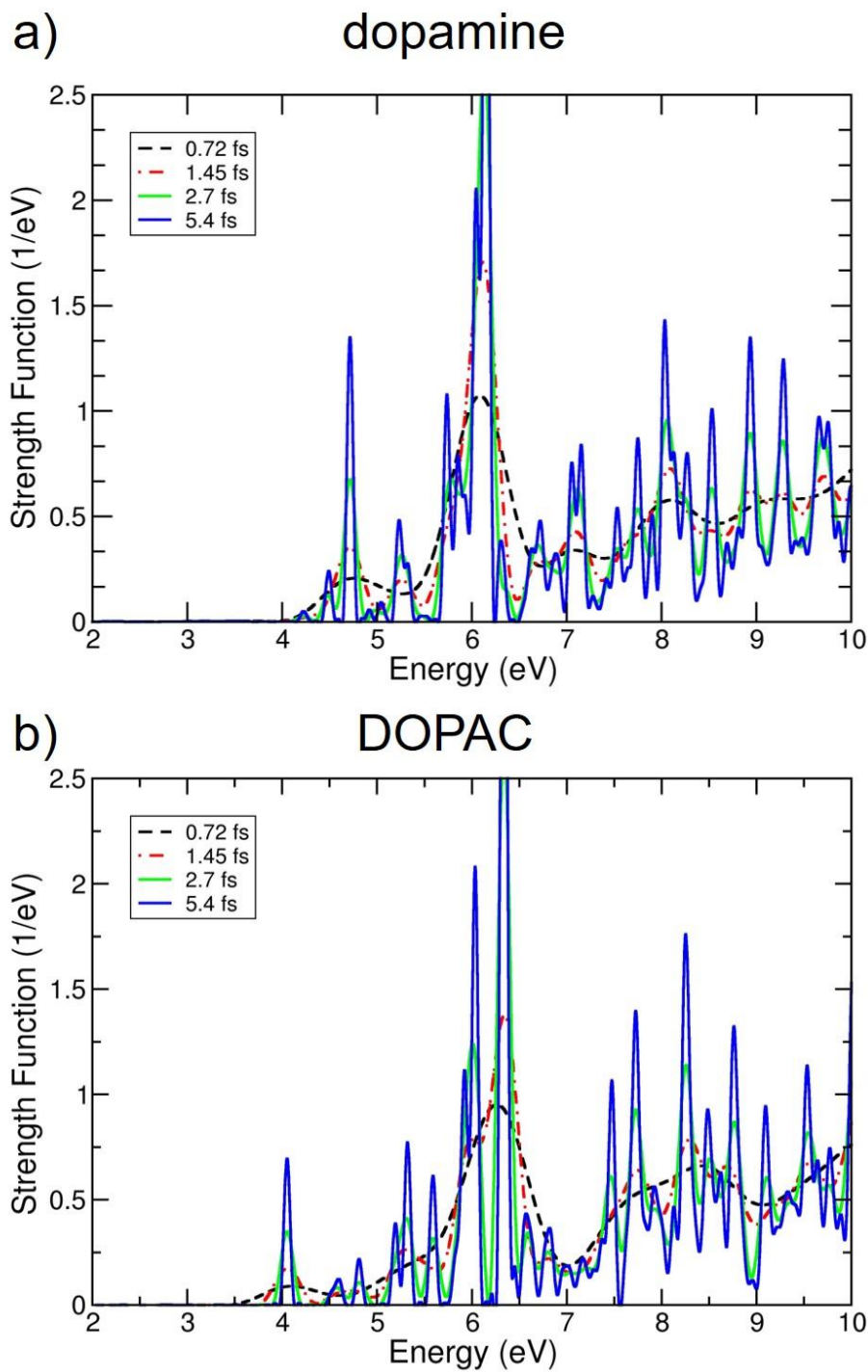

**Figure S7:** Calculated TDDFT absorption spectra for the isolated dopamine (a) and DOPAC (b) molecules with different propagation time which are indicated on the graphs. Sharper peaks are observed as the propagation time increases.

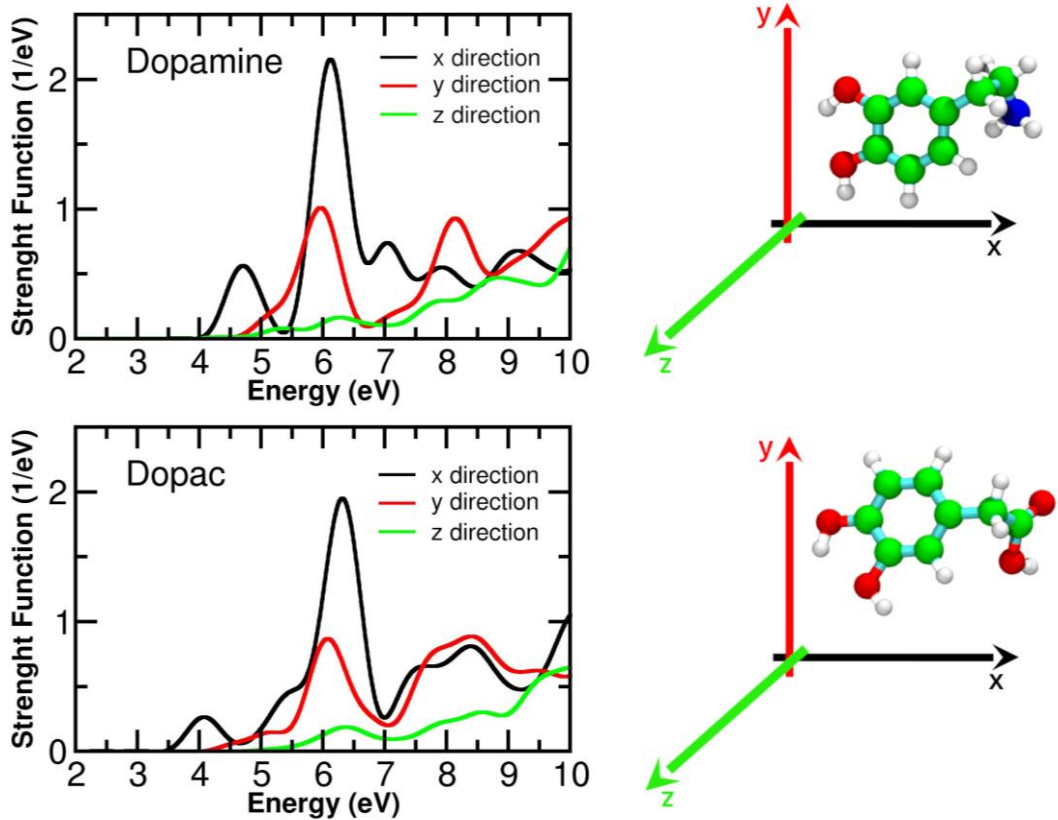

**Figure S8:** Three separate components of the strength function for light polarized along *x*,*y*, and *z*, according to the orientation of the centered molecule. *Top* Dopamine. *Down* DOPAC.

**Electronic structure from projected density of sates and charge density difference plot for a dopamine adsorbed on TiO₂ anatase (101) surface.**

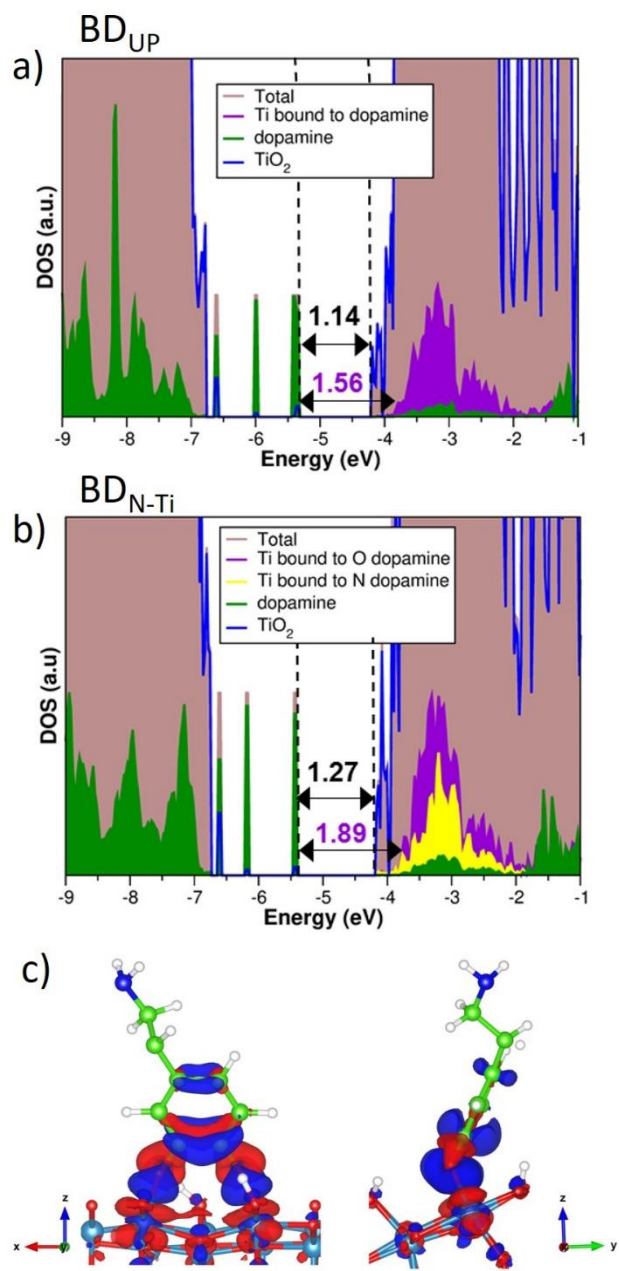

**Figure S9** Total (DOS) and projected (PDOS) density of states for a single molecule of dopamine adsorbed on the anatase (101) TiO₂ surface in **a)** BD$_{UP}$ and **b)** BD$_{N-Ti}$ configuration. The zero-energy reference is set at the vacuum level. The HOMO-CB gap in eV is reported in black, while the energy difference between the states of Ti directly connected to O of the molecule and the HOMO (ΔTi$_{bound}$-HOMO) is violet. c) Charge density difference plot for dopamine adsorbed on the anatase (101) surface. The red distribution corresponds to charge accumulation and blue correspond to depletion. The isosurface density value used is $5 \times 10^{-3}$ e/Å³.

**Table S3** Band gap calculated as a difference of Kohn-Sham eigenvalues obtained with the PBE functional between the molecular HOMO and the bottom of the CB ($E_g$), absorption onset ($A_{ons}$) and energy difference between the empty states of Ti bound to dopamine ($Ti_{bound}$) and the HOMO localized on the molecule ($\Delta Ti_{bound}$-HOMO) for the bare TiO$_2$ surface and for the the BD$_{UP}$ and the BD$_{Ni-Ti}$ configuration.

|  | **HOMO-CB (eV)** | $A_{ons}$ **(eV)** | **ΔTi$_{bound}$-HOMO (eV)** |
|---|---|---|---|
| **Bare TiO$_2$** | 2.54 | 2.30 | - |
| **BD$_{UP}$** | 1.14 | 1.50 | 1.56 |
| **BD$_{N-Ti}$** | 1.27 | 1.70 | 1.89 |